\documentclass[FBSedit,FBSmath,ecsub]{FBSsuppl}
\usepackage{amsfonts}
\usepackage{amssymb}
\usepackage{epsfig}
\newcommand{\be}{\begin{equation}}
\newcommand{\ee}{\end{equation}}
\newcommand{\ba}{\begin{eqnarray}}
\newcommand{\ea}{\end{eqnarray}}
\newcommand{\nsigma}{\mbox{\boldmath $\sigma$}}

\newcommand{\nl}{{\bf      l}}
\newcommand{\nn}{{\bf      n}}

\newcommand{\np}{{\bf      p}}       
\newcommand{\nq}{{\bf      q}}
         
\newcommand{\ns}{{\bf      s}}

\newcommand{\nP}{{\bf      P}}

\title{Relativity in polarized electron scattering observables}
\author{J.A. Caballero\instnr{1,}\thanks{\textit{E-mail address:} 
juan@nucle.us.es}, M.C. Mart\'{\i}nez\instnr{1}, T.W. Donnelly\instnr{2}, E. Moya de Guerra\instnr{3}, J.M. Ud\'{\i}as\instnr{4}, J.R. Vignote\instnr{4}}
\instlist{Departamento de F\'{\i}sica At\'omica, Molecular y Nuclear, Universidad de Sevilla,
E-41080 Sevilla, Spain
\and Center for Theoretical Physics, Massachusetts Institute of Technology, Cambridge, MA 02139, USA
\and Instituto de Estructura de la Materia, CSIC, E-28006 Madrid, Spain
\and Departamento de F\'{\i}sica At\'omica, Molecular y Nuclear, Universidad Complutense de Madrid, E-28040 Madrid, Spain}
\begin{document}

\maketitle
\begin{abstract}
Coincidence scattering of polarized electrons from nuclei with
polarization transfer to outgoing nucleons is studied within the context of relativistic
mean field theory. Effects introduced by the dynamical enhancement
of the lower components of the bound and scattered nucleon wave functions are analyzed
for the polarized response functions and transferred polarization asymmetries.
Results obtained
by projecting out the negative-energy components are compared with the
fully-relativistic calculation. The
crucial role played by the relativistic dynamics
in some spin-dependent observables is clearly manifested 
even for low/medium values of the missing momentum.
Kinematical relativistic effects are also analyzed.
A discussion of the factorization approach and the
mechanism for its breakdown is also briefly presented.

\end{abstract}

Quasielastic coincidence electron scattering reactions have provided over the years important 
insight into single-particle properties of nuclei. This is so because at
quasielastic kinematics the reaction mechanism underlying $(e,e'N)$
reactions can be treated with confidence in the impulse approximation (IA), i.e., assuming the 
virtual photon attached to a single bound nucleon that absorbs the whole momentum ($q$) and 
energy ($\omega$) (see~\cite{Bof96,Kel96} for details).

A large fraction of the theoretical analyses of $(e,e'N)$ reactions in past years was carried out on
the basis of non-relativistic calculations. Within this scheme, the bound and ejected nucleons are
described by non-relativistic wave functions which are solutions of the Schr\"odinger equation with
phenomenological potentials. Moreover, the current operator is also described by a non-relativistic 
expression derived directly from a Pauli reduction. Such non-relativistic reductions constitute the
basis for the standard distorted-wave impulse approximation (DWIA) that has been widely used to
describe $(e,e'N)$ experiments performed at intermediate energies~\cite{Bof96,Kel96}.

In the last decade some experiments performed have involved momenta and energies high enough to invalidate
the non-relativistic expansions assumed in DWIA. A consistent description of these processes requires 
one to incorporate relativistic degrees of freedom wherever possible.
Within this context,
nuclear responses and cross sections have been
investigated recently by our group using the relativistic mean field approach~\cite{Udi}. This
constitutes the basis of the relativistic distorted-wave impulse approximation (RDWIA), where bound
and scattered wave functions are described as Dirac solutions with scalar and vector potentials, and the
relativistic free nucleon current operator is assumed. 

Relativistic contributions can be cast into two general
categories, kinematical and dynamical relativistic effects. The former
are directly connected with the structure of the 
4-vector current operator, compared with the non-relativistic one that usually
involves $p/M_N$, $q/M_N$ and $\omega/M_N$
expansions. The latter, dynamical relativistic effects, come from the
difference between the relativistic and non-relativistic 
nucleon (bound and ejected) wave functions involved.
Within these dynamical relativistic effects one may distinguish
effects associated with the Darwin term (dynamical depression of the upper component of the
scattered nucleon wave function in the nuclear interior that
mainly affects the determination of spectroscopic factors at low 
missing momenta) and effects due to the dynamical enhancement of 
the lower components of the relativistic wave functions 
(expected to be especially relevant at high 
missing momenta, although they have proven to play an important role for
some particular observables even at low/medium $p$ values).

So far, fully-relativistic analyses of $(e,e'p)$ reactions have clearly 
improved the comparison with experimental data~\cite{Udi,Udi01}.
In recent work we have undertaken a systematic study of the 
relativistic effects in different observables. We started with
the relativistic plane-wave impulse approximation (RPWIA), 
i.e., neglecting final-state interactions (FSI) between the
outgoing nucleon and the residual nucleus. Although being an oversimplication, the RPWIA 
approach allows one to simplify the analysis,
disentangling the relativistic effects from other distortion effects.
The presence of negative-energy components in the 
relativistic bound nucleon wave function was shown to be very important
for some observables even at low/moderate values of the missing momentum.
In particular, the interference
$TL$ and $TT$ responses were proved to be very sensitive to
dynamical effects of relativity affecting the lower components.
These results persist also when FSI are included. 

Following similar arguments, we have presented in~\cite{Cris} a systematic study
of the new response functions that enter in the description of $A(\vec{e},e'\vec{N})B$ processes.
Since spin and relativity go hand in hand, one may {\it a priori} consider
the relativistic approach to be better suited to describe nucleon polarization
observables. Kinematical and dynamical relativistic effects for responses and polarization observables
have been analyzed in detail within the plane wave limit for the outgoing nucleon. In work in
progress effects in the final state are also being incorporated through relativistic FSI. They 
are briefly summarized in this work.


In the case of $A(\vec{e},e'\vec{N})B$ reactions, the hadronic response functions are
usually given by referring the recoil nucleon polarization to the
coordinate system defined by the
axes: $\nl$ (parallel to the momentum  $\np_N$ of the outgoing nucleon), $\nn$ (perpendicular to the plane 
containing $\np_N$ and the momentum transfer $\nq$), and $\ns$  (determined by
$\nn \times \nl$). In terms of the polarization asymmetries, the differential cross section can
be expressed in the form
\be
\frac{d\sigma}{d\varepsilon' d\Omega_ed\Omega_N}=
\frac{\sigma_0}{2}\left[1+\nP\cdot\nsigma+h\left(A+\nP'\cdot\nsigma\right)\right]\, ,
\label{eq11s3}
\ee
where $\sigma_0$ is the unpolarized cross section, $A$ denotes the
electron analyzing power, and $\nP$ ($\nP'$) represents the induced
(transferred) polarization. 
A general study of the properties and symmetries of all of these responses and
polarizations can be found in~\cite{Pick87}.
In the plane wave limit for the outgoing nucleon, the induced polarization $\nP$ and the
analyzing power $A$ are zero.
In terms of nuclear responses, from the total of eighteen response functions,
only nine survive within RPWIA. Four, $R^L_0$, $R^T_0$, $R^{TL}_0$ and $R^{TT}_0$, 
represent the unpolarized responses and
the five remaining, $R^{T'}_l$, $R^{T'}_s$, $R^{TL'}_l$, $R^{TL'}_s$ and $R^{TL'}_n$ (this one enters only
for out-of-plane kinematics), 
depend explicitly on the recoil nucleon polarization and only enter when the 
electron beam is also polarized. In this work, we restrict our discussion to these observables 
known as transferred polarization responses and/or transferred asymmetries.

First, we consider the case of the plane wave limit for the outgoing nucleon wave function (RPWIA)
The role played by the negative-energy projection components in the polarized
responses was discussed in~\cite{Cris}. There we show
that in two responses, $R^{T'}_l$ and $R^{TL'}_s$, the contribution 
of the negative-energy projections is almost negligible, that is, dynamical relativistic effects 
from the bound nucleon wave function do not significantly affect these responses. 
On the contrary, $R^{T'}_s$ and $R^{TL'}_l$ are much more sensitive.
This result resembles what appeared for the unpolarized interference
$TL$ response~\cite{Cab98}. Hence there exists a strong discrepancy 
between RPWIA results and those corresponding to the standard factorized PWIA.

\begin{figure}
{\par\centering \resizebox*{0.75\textwidth}{0.35\textheight}{\rotatebox{270}
{\includegraphics{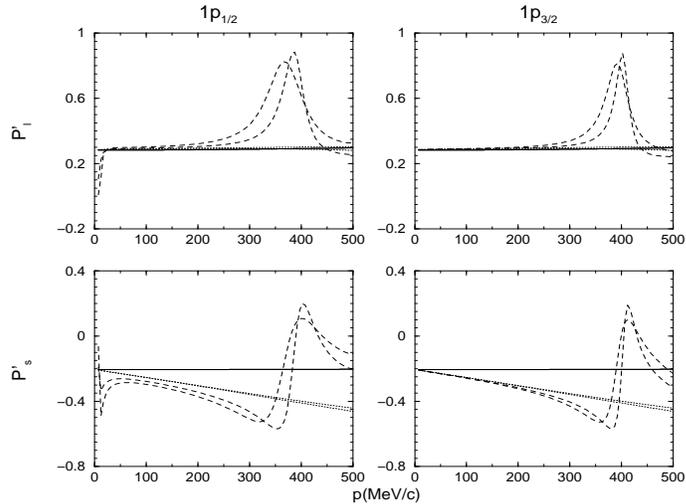}}} \par}
\caption{\label{sec33fig3} Transferred polarization asymmetries $P'_l$ and $P'_s$
in the plane wave limit for the outgoing nucleon.
Fully-relativistic results (dashed lines)
are compared with their positive-energy projection contributions (dotted lines). Thin
lines correspond to the $CC1$ current operator and thick lines to $CC2$. 
We also show for comparison the static limit result (solid line).}
\end{figure}

Dynamical enhancement of the lower components in the bound nucleon wave
function is even more clearly seen when analyzing the transferred polarization asymmetries
(Fig.~\ref{sec33fig3}). Here 
the fully-relativistic RPWIA results (dashed lines) corresponding to the Coulomb gauge with
the $CC1$ and $CC2$ choices of the current (see ref.~\cite{Cris} for details on the
current operators), are compared with the transferred polarizations
obtained by projecting out the negative-energy components (dotted lines).
Results for $p_{1/2}$ (left-hand panels) and $p_{3/2}$ (right-hand panels) are shown. Kinematics 
corresponds to $q=1$ GeV/c, $\omega=440$ MeV/c and forward ($\theta_e=23^0$) scattering
angle. First, note the difference between the relativistic 
and projected results observed at very small missing momentum values for the $p_{1/2}$ shell.
This effect comes directly from the quantum number $\overline{\ell}$ involved in the
lower component of the bound state wave function ($\overline{\ell}=0$ for $p_{1/2}$).
Moreover, it is also important to point out that
fully-relativistic and positive-energy projected
results typically do not differ appreciably
for $p$-values up to $\sim 300$ MeV/c. For $p > 300$ MeV/c
relativistic and projected results start to deviate from each
other. This general behaviour is what one expects because of
the clear dominance of the positive-energy projection component of the momentum
distribution in the region $p\leq 300$ MeV/c~\cite{Cab98}.
On the contrary, for $p>300$ MeV/c the negative-energy components are
similar to or even larger than the positive ones, and hence the effects of the
dynamical enhancement of the lower components in the bound
relativistic wave functions are clearly visible in the transfer polarization asymmetries.

\begin{figure}
{\par\centering \resizebox*{0.85\textwidth}{0.5\textheight}{\rotatebox{270}
{\includegraphics{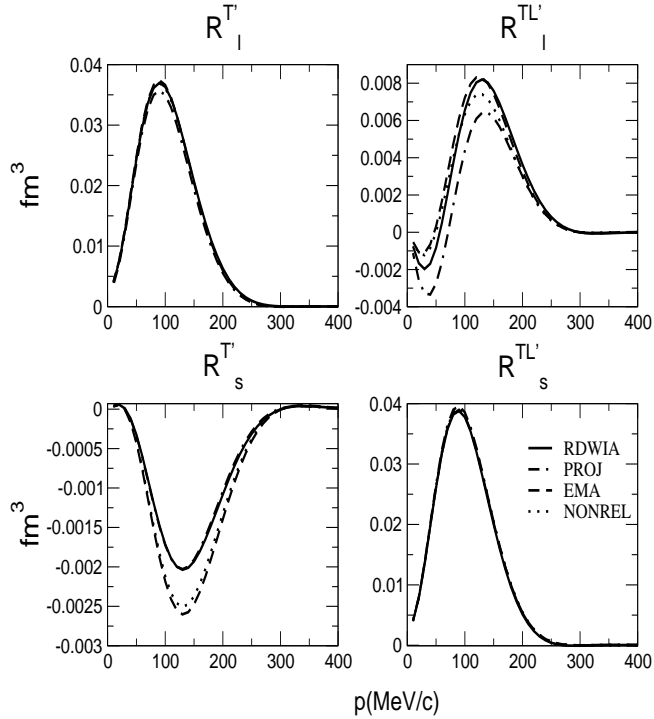}}} \par}
\caption{\label{fig2}Polarized response functions for the \protect\( 1p_{1/2}\protect \)
shell. Fully-relativistic
response (solid line) is compared with the projected one (dot-dashed), 
EMA (dashed) and non-relativistic approach (dotted).}
\end{figure}

Obviously, final state interactions (FSI) between the ejected nucleon and the residual nucleus 
should be included in the analysis of $A(\vec{e},e'\vec{N})$ processes in order to compare with
data. In fact, FSI introduce significant modifications in the responses and transferred polarization
asymmetries. However, the high sensitivity of polarization-related-observables to negative-energy
projections shown within RPWIA is maintained in the relativistic
distorted-wave impulse approximation (RDWIA). This was already the case for the unpolarized interference
longitudinal-transverse response $R^{TL}$ and $A_{TL}$-asymmetry. 
The analysis presented in~\cite{Udi99} proves the crucial role played
by both kinematical and dynamical relativistic effects in order to fit the experiment. In particular,
the richness shown by the structure of the left-right asymmetry ($A_{TL}$) is only consistent with
predictions of relativistic calculations that include dynamic enhancement of the lower components of
Dirac spinors. 

Similar comments can be also applied to the transferred polarization observables. In particular, 
the different sensitivity to relativistic effects shown by the transferred polarized responses
within RPWIA, persists when RDWIA calculations are involved. This
is illustrated in Fig.~2 where we present the
polarized responses corresponding to the $p_{1/2}$-shell in $^{16}$O. Kinematics 
has been selected as in the previous
figure. We compare the fully distorted relativistic calculation using the Coulomb gauge and the
current operator CC2 (solid line) with the results after projecting the bound and scattered proton
wave functions over positive-energy states (dot-dashed) and using the asymptotic momenta (dashed). 
This last approach is equivalent to the effective momentum approximation (EMA).
Also for comparison we show a non-relativistic calculation where the non-relativistic current
operator (obtained from a Pauli reduction) corresponds to the expression given in~\cite{Amaro}.
Note that two responses, $R^{TL'}_s$ and $R^{T'}_l$, are very insensitive to dynamical and kinematical
relativistic effects. On the contrary, effects due to relativity are clearly visible for
$R^{TL'}_l$ and $R^{T'}_s$. This result agrees with the findings within RPWIA~\cite{Cris}.

\begin{figure}
{\par\centering \resizebox*{0.7\textwidth}{0.5\textheight}{\rotatebox{270}
{\includegraphics{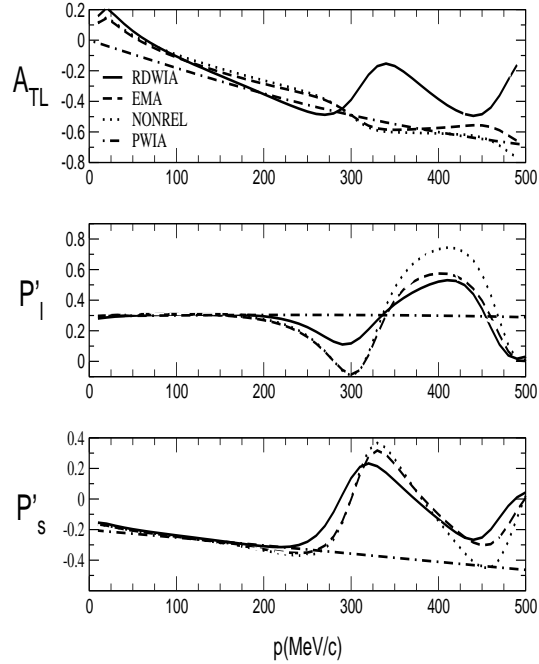}}} \par}
\caption{\label{fig3}Asymmetry $A_{TL}$ and transferred polarization ratios $P'_l$ and $P'_s$. RDWIA 
results (solid) are compared with EMA (dashed), non-relativistic (dotted) and factorized PWIA (dot-dashed)
calculations.}
\end{figure}

Finally, in Fig.~3 we show the RDWIA results for the left-right asymmetry $A_{TL}$ and transferred
polarization asymmetries $P'_l$ and $P'_s$. The labels of the different curves are
shown in the caption. It is
particularly interesting the comparison between the fully relativistic results 
and the ones corresponding
to the asymptotic projection (equivalent to EMA) and
the non-relativistic calculation. Note that the richer structure 
shown by the relativistic calculation in
$A_{TL}$ is basically lost within the EMA and non-relativistic approaches. Moreover, these results get
closer to the factorized PWIA calculation. This modest deviation from the factorized (free) result is
mainly due to the spin-orbit coupling in the final state. The behaviour shown by the transferred
polarization asymmetries is clearly different. Here, the EMA and non-relativistic calculations do no
change substantially the structure shown by the fully relativistic result,
differing clearly from
the factorized (free) one. This means that the spin-orbit coupling in
the final wave function
breaks down completely factorization in the polarized observables.

Summarizing, in this work we have analyzed outgoing nucleon polarized observables within a
relativistic mean field approach. We have shown that dynamical relativistic effects, namely the
enhancement of the lower components of Dirac wave functions, 
affect very significantly the observables. This was already the
case for the unpolarized observables, $R^{TL}$ and $A_{TL}$. However, here we also show that
the behaviour presented by the transferred polarization asymmetries is clearly different from the one
corresponding to the left-right asymmetry.

{\it Acknowledgements.} This work was supported by funds provided by DGICYT (Spain) 
under Contracts Nos. PB/98-1111, PB/98-0676 and the Junta
de Andaluc\'{\i}a (Spain) and by the U.S. Department of Energy under
Cooperative Research Agreement No. DE-FC02-94ER40818.
M.C.M acknowledges support from a fellowship from the Fundaci\'on
C\'amara (University of Sevilla). J.A.C. also acknowledges MEC (Spain) for a sabbatical stay at
MIT (PR2001-0185).


\end{document}